\documentclass[a4paper]{article}

\usepackage{INTERSPEECH2021}
\usepackage{array}
\usepackage{microtype}
\usepackage{url}
\usepackage{appendix}

\newcolumntype{x}[1]{%
>{\centering\hspace{0pt}}p{#1}}%
\newcommand{\tnhl}{\tabularnewline\hline}

\title{GISE-51: A scalable isolated sound events dataset}
\name{Sarthak Yadav, Mary Ellen Foster}
\address{
  University of Glasgow, UK
}
\email{s.yadav.2@research.gla.ac.uk, MaryEllen.Foster@glasgow.ac.uk}

\begin{document}

\maketitle
\begin{abstract}
Most of the existing isolated sound event datasets consist of a small number of sound event classes, usually 10 to 15, restricted to a small domain, such as domestic and urban sound events. In this work, we introduce GISE-51, a dataset spanning 51 isolated sound events belonging to a broad domain of event types, derived from FSD50K \cite{fonseca2020fsd50k}. We also release GISE-51-Mixtures, a dataset of 5-second soundscapes with hard-labelled event boundaries synthesized from GISE-51 isolated sound events. We conduct baseline sound event recognition (SER) experiments on the GISE-51-Mixtures dataset, benchmarking prominent convolutional neural networks, and models trained with the dataset demonstrate strong transfer learning performance on existing audio recognition benchmarks. GISE-51 provides freedom to adapt the included isolated sound events for domain-specific applications.
\end{abstract}

\section{Introduction}
Most of the sound event datasets can be divided into two main categories: (i) weakly-labelled datasets with instance-level tags with labels spanning a large hierarchy of sound events at various granularities \cite{audioset,Fonseca2019audio,chen2020vggsound,fonseca2020fsd50k}, and (ii) hard-labelled datasets \cite{turpault2019sound} with frame-level labels and additional metadata. Due to their increased curation complexity, hard-labelled datasets often consist of a small number of samples spanning a small event vocabulary limited to a single domain. As a result, they are often paired with weakly-labelled and synthesized sound mixtures data for model development \cite{Turpault2019, Adavanne2018_JSTSP}. In contrast, weakly-labelled datasets tradeoff hard-labelling for sheer dataset size and real-world acoustic content coverage.

An alternative strategy is to synthesize sound event mixtures from isolated, single-source event instances \cite{turpault2019sound, salamon2017scaper, fuss}. Paired with various background/environmental noises and room impulse-response augmentation, this allows for automated generation of a large number of soundscapes with significantly lower label noise and accurate event metadata. However, isolating sound events from real-world data also requires significant curation efforts, and most of the currently available isolated event corpora feature a restricted vocabulary, focusing on a few classes in a narrow domain \cite{turpault2019sound, salamon2017scaper}.

In this work, we introduce \textit{Glasgow Isolated Sound Events} (GISE-51)\footnote{doi.org/10.5281/zenodo.4593514}, a dataset of isolated sound events spanning 51 classes. GISE-51 is derived from the recent FSD50K dataset \cite{fonseca2020fsd50k}, which is a CC-licensed release based on audio clips from \textit{freesound.org} \cite{fonseca2017freesound}. We also propose \textit{GISE-51-Mixtures}, a dataset of 5-second soundscapes with up to three sound events synthesized from GISE-51. GISE-51 provides the freedom to adapt the provided isolated sound events for domain-specific applications by pairing with various backgrounds/environments.\footnote{https://github.com/SarthakYadav/GISE-51-pytorch}

\section{Related work}
Several sound event datasets have been proposed recently. AudioSet \cite{audioset} is the largest publically available sound event dataset, with $\approx$2.1 M audio clips spanning an ontology of 527 sound event classes, with multiple event labels per utterance. However, AudioSet is based on audio clips taken from YouTube videos, which are not a part of the official release. This makes acquiring the dataset and reproducing previous results complicated since YouTube video availability varies with geography and is subject to removal over time. VGGSound \cite{chen2020vggsound} is a more recent large-scale dataset spanning 309 classes generated using automated computer vision techniques. It is also based on YouTube videos and inherits the same shortcomings as AudioSet. 

FSD50K \cite{fonseca2020fsd50k} addresses the aforementioned shortcomings by releasing a manually annotated dataset of multi-labelled audio clips spanning 200 classes, facilitating ease of acquisition without missing labels at the instance level. 
For the proposed GISE-51 dataset, we process the multilabel annotations in FSD50K to obtain isolated, single event instances belonging to 51 classes, and the synthetic soundscapes generated using the processed isolated instances are of homogenous duration with well-defined event boundaries, unlike FSD50K, which has audio clips of variable duration.

Most of the isolated sound event-based synthetic datasets span very few event classes. URBAN-SED \cite{salamon2017scaper} is a dataset of 10,000 ten-second synthetic soundscapes generated from isolated sound event clips from UrbanSound8k \cite{salamon2014dataset}, spanning ten urban sound event classes. The DESED dataset \cite{Turpault2019}, used in the DCASE 2020 Task 4 challenge, consists of domestic recordings and synthesized mixtures spanning ten sound events, the majority of which correspond to domestic appliances (Alarm/bell/ringing, Blender, Dishes, Electric shaver/toothbrush, Vacuum cleaner). A similar trend can be observed across several iterations of DCASE challenges in the past \cite{Turpault2019,Serizel2018}. In contrast, GISE-51 covers a significantly greater number of sound events from a larger domain, ranging from vehicle sounds, environmental sounds, musical instruments and human speech.

\section{GISE-51 and GISE-51-Mixtures datasets}

We refer to the dataset of isolated events as the GISE-51 dataset and the corresponding mixtures as the GISE-51-Mixtures. The following sections cover the two proposed datasets in more detail.

\subsection{GISE-51 Dataset}

The GISE-51 dataset consists of 16,357 clips of variable duration, with each clip containing a single sound event out of a 51 class vocabulary derived from FSD50K \cite{fonseca2020fsd50k}. The dataset is divided into three subsets: \textit{train}, \textit{val} and \textit{eval} with 12465, 1716 and 2176 utterances, respectively. The subsets are congruent with the original FSD50K release \cite{fonseca2020fsd50k}---that is, files in a given subset in FSD50K (say \textit{train}) belong to the same subset in GISE-51. The following process yields isolated sound events spanning 51 classes. However, despite our best efforts, some label noise is expected to remain as we did not manually annotate all the utterances.

\subsubsection{Unsmearing annotations}
The FSD50K ontology consists of 144 leaf nodes and 56 root nodes, and the ground truth annotations have smeared labels, i.e. ``the labels are propagated in the upwards direction to the root of the ontology". The FSD50K release provides both raw collection data with unsmeared labels and the final, processed and verified smeared ground truth data. To obtain utterances with a single sound event we proceed as follows: (i) select those utterance files which have a single raw collection label, removing other files; (ii) cross-reference the presence of this label in the final smeared ground truth, shortlisting utterances that satisfy this condition, resulting in around 20k utterances. 
The isolated sound event clips are then split into \textit{train, val} and \textit{eval} subsets following the FSD50K release.

\subsubsection{Manual Inspection}

We randomly sampled 100 (or 20\%, whichever is lower) utterances from each sound event class for manual inspection for qualitative evaluation. The objectives of this step were: (i) to determine the fraction of utterances with more than one event; (ii) to determine the fraction of utterances with ambiguous labels (especially with sound events in the musical instruments domain); and (iii) to determine classes with the greatest inter-class ambiguity and which of these classes could be merged.

\subsubsection{Finalizing vocabulary}
At this stage, the remaining sound event classes are re-evaluated to be kept as is, merged or discarded. Sound events that were not flagged for merging and had too few samples ($<$75) were removed (e.g. \textit{Mechanical Fan}). Some sound event classes were dropped to maintain overall diversity in the dataset because sufficient classes of that type were already present in the dataset. Table~\ref{tab:merged_classes} lists the merged classes. There are several reasons to merge these labels, including: (i) to undo the effects of \textit{unsmearing} labels, which can increase ambiguity across root-leaf labels and semantically similar events (ii) to improve statistics of similar/related sound events which, independently, had insufficient samples; (iii) to merge ambiguous sound events flagged during the manual inspection. Finally, after filtering classes with very few instances, we are left with 51 sound event classes. (Appendix \ref{appendix_classes})

\subsubsection{Silence Filtering}
We employed \textit{sox}\footnote{\url{http://sox.sourceforge.net/}} for volume threshold-based automatic silence filtering. Based on trial-and-error, sound event classes are divided into bins with different volume thresholds (more details on the dataset page).

\begin{table}[th]
  \caption{List of merged sound event classes}
  \label{tab:merged_classes}
  \centering
  \begin{tabular}{x{0.7\linewidth}}
    \toprule
    \textbf{Classes Merged}\tnhl
    \midrule
    Acoustic\_guitar, Bass\_guitar, Guitar \tnhl
    Bass\_drum, Snare\_drum \tnhl
    Child\_speech\_and\_kid\_speaking, Female\_speech\_and\_woman\_speaking, Male\_speech\_and\_man\_speaking \tnhl
    Cutlery\_and\_silverware,
    Dishes\_and\_pots\_and\_pans \tnhl
    Bark, Dog \tnhl
    Crumpling\_and\_crinkling, Crushing \tnhl
    Waves\_and\_surf, Splash\_and\_splatter \tabularnewline

    \bottomrule
  \end{tabular}
\end{table}


\subsection{GISE-51-Mixtures Dataset}

The GISE-51-Mixtures subset is created using Scaper \cite{salamon2017scaper}, a tool for automated soundscape generation. Scaper has several tunable sound event parameters that can be sampled from user-defined distributions, allowing the creation of randomized soundscapes from target background and foreground audio clips. For generating synthetic 5-second soundscapes, the background is sampled at random from typical noise colours. Each soundscape can have up to 3 foreground events, which are random segments selected from an isolated sound event file. Weighted sampling with replacement is used for selecting the isolated event file, effectively oversampling sound event classes with fewer utterances in the resulting soundscapes. The mixtures data is divided into \textit{train}, \textit{val} and \textit{eval} subsets generated from the corresponding subsets of the isolated events data, with 60000, 10000 and 10000 soundscapes in each subset, respectively. The actual mixtures and the code for generating these mixtures from annotation files and isolated sound events are provided as a part of the release.

\section{Experiments}
The following section describes the experiments conducted on the GISE-51 Mixtures dataset. Previously addressed using classical approaches such as those based on i-vectors \cite{huang2013blind}, Mel-frequency cepstrum coefficient (MFCC) features and Gaussian Mixture models \cite{zhuang2010real}, there has been a recent shift towards applying deep neural networks for sound event recognition, including CNNs \cite{chen2020vggsound, hershey2017cnn, Akiyama2019, kong2020panns, Du2020_task3_report}, Recurrent Neural Networks (RNNs) \cite{parascandolo2016recurrent, xu2018large}, as well as CNN-RNN hybrid approaches \cite{Ebbers2019}. Instead of proposing a custom architecture, we instead focus on benchmarking convolutional neural networks from the prominent ResNet \cite{he2016deep}, DenseNet \cite{huang2017densely} and EfficientNet \cite{tan2019efficientnet} paradigms to serve as baselines for future research. We further compare the transfer learning performance of CNNs pre-trained on the GISE-51-Mixtures dataset on the AudioSet balanced track, the VGGSound dataset and the ESC-50 dataset.

All experiments use Adam optimiser with a default learning rate of 1e-3, unless stated otherwise, using zero-centred log-spectrogram features calculated with a window of 20 ms and a stride of 10 ms from 22050 Hz audio files. Most of the experiments are done with a per-GPU-process batch size of 64. Models are trained for a maximum of 50 epochs, with learning rate reduced by a factor of 10 when no improvement is observed for five epochs, and training is stopped early if no improvement is observed for ten epochs. Following previous works \cite{chen2020vggsound, fonseca2020fsd50k, hershey2017cnn}, we adopt \textit{mean average precision (mAP)} and \textit{d-prime} as our primary performance metrics for experiments conducted on GIST-51-Mixtures, AudioSet ``balanced" and VGGSound datasets.

\subsection{Number of synthesized soundscapes v/s \textit{val} mAP}

\begin{figure}[t]
  \centering
  \includegraphics[width=\linewidth]{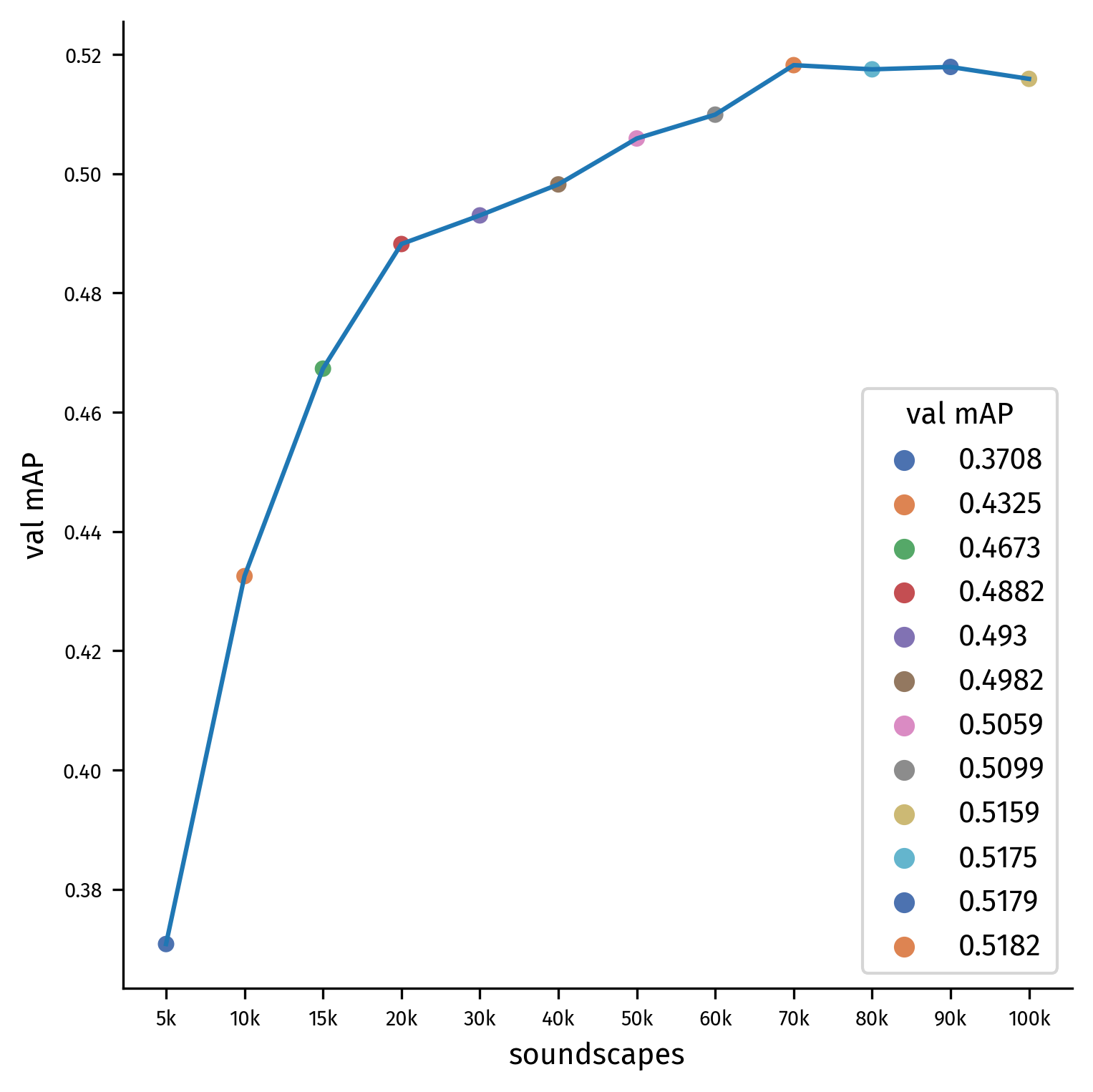}
  \caption{Number of soundscapes v/s val mAP}
  \label{fig:mixtures_vs_mAP}
\end{figure}

Although we can synthesize any arbitrary number of soundscapes, due to the limited number of isolated sound event utterances, oversampling the same sound events repeatedly might not necessarily correspond to better generalization performance. To this end, we study how the number of training soundscapes affects \textit{val} set (which is kept fixed) performance by training ResNet-18 \cite{he2016deep} on varying number of soundscapes in the $[5000, 100000]$ range. Each experiment is repeated at least three times, and average results are reported.
From Figure~\ref{fig:mixtures_vs_mAP}, we can observe that initially, \textit{val} mAP improves drastically with the number of soundscapes, gradually tapering off and maxing out at the 70k mark. 60k training soundscapes provide a fair trade-off between performance and training time and is thus used in further experiments.

\subsection{GIST-51-Mixtures CNN Baselines}
We compare the performance of convolutional neural networks of various complexities trained on 60k synthetic soundscapes. Except EfficientNet-B0, EfficientNet-B1 \cite{tan2019efficientnet}, DenseNet-121 \cite{huang2017densely} and ResNet-50 \cite{he2016deep}, which were trained on a single Nvidia V100, all CNNs were trained on a 4x Nvidia 1080 machine. Mixup augmentation \cite{zhang2017mixup} was used during training as we observed it gave a significant boost in performance (for the ResNet-18 model, it improved \textit{eval} mAP score by $\approx$0.03). All experiments were repeated at least three times, and average metrics are reported. 
Table~\ref{tab:mixtures_res} reports the \textit{mAP} and \textit{d-prime} scores on the \textit{eval} set for each model. ResNet-18 and EfficientNet-B1 are the worst and the best performing models, respectively. Note that ResNet-44 and ResNet-56 are the variants proposed in \cite{he2016deep} for the CIFAR-10 and CIFAR-100 datasets \cite{krizhevsky2009learning}, and both perform remarkably well as compared to other larger networks from the ResNet family, possibly due to over-parameterisation in the latter.

For further experiments, we include only the worst- and best-performing models (ResNet-18 and EfficientNet-B1), as one can extrapolate the results of other architectures based on the results on these models.

\begin{table}[th]
    \caption{\textit{Eval} performance on GISE-51-Mixtures dataset}
    \label{tab:mixtures_res}
    \centering
    \begin{tabular}{c|c|c|c}
        \toprule
        \textbf{Model} & \textbf{Params} & \textbf{mAP} & \textbf{d-prime} \\
        \midrule
        ResNet-44* & 2.632 M & 0.5656 & 1.8293\\
        ResNet-56* & 3.408 M & 0.5634 & 1.8330\\
        \underline{ResNet-18} & \underline{11.196 M} & \underline{0.5551} & \underline{1.7956}\\
        ResNet-34 & 21.305 M & 0.5722 & 1.8472\\
        ResNet-50 & 23.606 M & 0.5677 & 1.8220\\
        \hline
        EfficientNet-B0 & 4.072 M & 0.5984 & 1.9155\\
        \textbf{EfficientNet-B1} & \textbf{6.578 M} & \textbf{0.6062} & \textbf{1.9303}\\
        \hline
        DenseNet-121 & 7 M & 0.6053 & 1.8640\\
        \bottomrule
    \end{tabular}
\end{table}

\subsection{Transfer learning on prominent benchmark datasets}
\subsubsection{AudioSet Balanced}
As previously stated, due to varying YouTube video availability, acquiring and comparing results on the AudioSet dataset is not straightforward. However, as AudioSet is the largest, most comprehensive sound event dataset, it is still a valuable benchmark. We evaluate ResNet-18 and EfficientNet-B1 models with and without GISE-51-Mixtures pretraining on the AudioSet balanced track, training on the AudioSet ``balanced train" subset. Table~\ref{tab:audioset_res} reports results on the AudioSet evaluation data. Several videos were unavailable; we were only able to acquire 19253 and 17498 videos for the ``balanced train" and ``evaluation" subsets, respectively.\footnote{The code release includes the list of YouTube video ids that were used.} For the transfer learning experiment, the best performing run of the corresponding GISE-51-Mixtures model was used. Instead of binary cross-entropy, we optimized Focal Loss \cite{lin2017focal} for training the model, along with Mixup augmentation. Each experiment was repeated at least three times, and the average metrics on the Evaluation set are reported. 

\begin{table}[th]
    \caption{AudioSet \textit{Evaluation} performance}
    \label{tab:audioset_res}
    \centering
    \begin{tabular}{c|c|c|c}
        \toprule
        \textbf{Model} & \textbf{Pre-trained} & \textbf{mAP} & \textbf{d-prime}\\
        \midrule
        ResNet-18 & False & 0.2053 & 2.1313\\
        EfficientNet-B1 & False & 0.2287 & 2.1392\\
        ResNet-18 & True & 0.2236 & 2.0673\\
        EfficientNet-B1 & True & 0.2595 & 2.1886\\
        \bottomrule
    \end{tabular}
\end{table}

\subsubsection{VGGSound}
VGGSound \cite{chen2020vggsound} is another recent large-vocabulary sound event recognition dataset collected and annotated via an automated computer vision pipeline. It is the largest dataset covered in our experiments, with around 300 classes and 200,000 utterances. Like \cite{chen2020vggsound}, we randomly sample a 5-sec audio clip from the 10-sec file while training, and at test time, perform inference on the entire clip. All models are trained on the \textit{VGGSound-train} set and results are reported on the \textit{VGGSound-test} set. As opposed to previous experiments, we optimize cross-entropy loss to train the model, since VGGSound has a single label per utterance. For ResNet-18, we only conduct transfer learning experiments (denoted by \textit{ft} in Table~\ref{tab:vggsound_res}) since from-scratch performance has already been documented \cite{chen2020vggsound}, whereas EfficientNet-B1 is trained and evaluated both from scratch and in a transfer learning setting. No hyperparameter tuning was performed, and both the models were trained at a batch size of 64. From Table \ref{tab:vggsound_res}, we observe that in comparison to the AudioSet balanced task, despite having a smaller vocabulary and an easier to optimize learning objective (multiclass setting v/s multilabel for AudioSet), transfer learning with GISE-51-Mixtures provides a much smaller improvement over training from scratch for the VGGSound dataset, observing a $\approx$1\% improvement in Top-1 Accuracy and \textit{mAP} scores for both the models. We conjecture that this is primarily due to dataset size, since VGGSound has approximately $10\times$ the number of utterances as the AudioSet balanced task.

\begin{table}[th]
    \caption{VGGSound \textit{Test} performance}
    \label{tab:vggsound_res}
    \centering
    \begin{tabular}{c|c|c|c}
        \toprule
        \textbf{Model} & \textbf{Top-1 Acc (\%)} & \textbf{mAP} & \textbf{d-prime} \\
        \midrule
        ResNet-18 \cite{chen2020vggsound} & 48.8 & 0.516 & 2.627 \\
        EfficientNet-B1 & 50.0 & 0.519 & 2.695\\
        ResNet-18 + ft & 49.9 & 0.520 & 2.707\\
        EfficientNet-B1 + ft  & 51.2 & 0.533 & 2.737\\
        \bottomrule
    \end{tabular}
\end{table}

\subsubsection{ESC-50}
ESC-50 \cite{piczak2015esc} is a recently released dataset of 2000 5-second long environmental sound recordings, organised into 50 semantic classes. ESC-50 adopts a 5-fold cross-validation protocol for evaluation. We perform transfer learning experiments on ResNet-18 and EfficientNet-B1 models, randomly sampling a 3-second clip followed by mixup augmentation while training. Cross-entropy loss with a weight decay of 1e-4 is minimised for training the models. Each model is trained on all folds three times, and average accuracy across runs is reported. As observable from Table \ref{tab:esc50_res}, the models pretrained on the proposed dataset perform well in comparison to several recent results, including ensemble models \cite{nanni2020ensemble, kimurban}, models that utilize additional metadata \cite{kimurban} as well as comprehensive multi-stage self-taught models trained on the entire AudioSet ontology \cite{kumar2020sequential}.

\begin{table}[th]
    \caption{ESC-50 evaluation}
    \label{tab:esc50_res}
    \centering
    \begin{tabular}{c|c}
        \toprule
        \textbf{Model} & \textbf{Acc\%} \\
        \midrule
        Human \cite{piczak2015esc} & 81.3 \\
        Aclnet (16 kHz) \cite{huang2018aclnet} & 80.9\\
        Aclnet (44.1 kHz) \cite{huang2018aclnet} & 85.65\\
        CNN Ensemble \cite{nanni2020ensemble} & 88.65\\
        EfficientNet ensemble \cite{kimurban} & 89.5\\
        WEANET \cite{kumar2020sequential} & 94.1\\
        \hline
        ResNet-18+ft \textbf{(ours)} & 83.92\\
        EfficientNet-B1+ft \textbf{(ours)}  & 85.72\\
        \bottomrule
    \end{tabular}
\end{table}

\section{Conclusion}
We have introduced GISE-51, a dataset of isolated sound events spanning 51 classes based on the FSD50K dataset \cite{fonseca2020fsd50k}, along with GISE-51-Mixtures, a dataset of synthetic soundscapes generated from GISE-51. Collectively, the GISE-51 release offers freedom for domain-specific adaptation along with providing an open, reproducible benchmark for system development. Several experiments are conducted to establish the baseline performance of prominent convolutional neural network architectures. Models trained on the proposed dataset demonstrate strong transfer learning performance on several audio recognition tasks, demonstrating the utility of the proposed dataset. The results also support the viability of the semi-synthetic dataset paradigm: curating isolated sound events and utilizing their mixtures instead of directly curating complex real-world soundscapes.

\section{Acknowledgement}
We thank Eduardo Fonseca of \cite{fonseca2020fsd50k}, for their valuable feedback and continued discourse. They provided experimental outcomes supporting notions contrary to some statements made in the previous version of this article, which we independently and concurrently observed in our later research.

\appendixtitleon
\appendixtitletocon
\begin{appendices}
\section{GISE-51 sound event classes}
\label{appendix_classes}
Applause, Boom and explosion, Burping and eructation, Bus, Car, Chirp and tweet, Clapping, Coin (dropping), Cough, Crash cymbal, Cricket, Crumpling and crinkling and crushing, Dishes and pots and pans and cutlery and silverware, Dog, Drill, Drum, Fart, Fireworks, Glass, Gong, Guitar, Gunshot and gunfire, Hammer, Harmonica, Harp, Hi-hat, Human speech, Keys jangling, Laughter, Marimba and xylophone, Organ, Piano, Rattle (instrument), Screaming, Shatter, Slam, Sliding door, Subway and metro and underground, Tambourine, Tap, Tearing, Thunder, Toilet flush, Train, Trumpet, Walk and footsteps, Waves and surf and splash and splatter stream, Whispering, Wind, Writing, Zipper (clothing)
\end{appendices}

\newpage
\bibliographystyle{IEEEtran}
\bibliography{mybib}

\end{document}